# MINIMALIST: switched-capacitor circuits for efficient in-memory computation of gated recurrent units

Sebastian Billaudelle⋆ @, Laura Kriener⋆, Filippo Moro, Tristan Torchet, Melika Payvand
Institute of Neuroinformatics, University of Zürich and ETH Zürich
⋆ contributed equally @ sebastian@ini.uzh.ch

*This preprint represents an early and preliminary version of the final manuscript and will be updated regularly.*

*Recurrent neural networks (RNNs) have been a long-standing candidate for processing of temporal sequence data, especially in memory-constrained systems that one may find in embedded edge computing environments. Recent advances in training paradigms have now inspired new generations of efficient RNNs. We introduce a streamlined and hardware-compatible architecture based on minimal gated recurrent units (GRUs), and an accompanying efficient mixed-signal hardware implementation of the model. The proposed design leverages switched-capacitor circuits not only for in-memory-computing (IMC), but also for the gated state updates. The mixed-signal cores rely solely on commodity circuits consisting of metal capacitors, transmission gates, and a clocked comparator, thus greatly facilitating scaling and transfer to other technology nodes.*

*We benchmark the performance of our architecture on time series data, introducing all constraints required for a direct mapping to the hardware system. The direct compatibility is verified in mixed-signal simulations, reproducing data recorded from the software-only network model.*

## 1 Introduction

Modeling temporal signals is a core challenge in AI, with applications ranging from speech and language to sensory processing and control. RNNs, particularly gated variants like long short-term memories (LSTMs) (Hochreiter 1997) and GRUs (Cho et al. 2014), emerged as the natural choice for such tasks due to their ability to maintain an internal memory and process inputs sequentially. They offered a principled way to capture temporal dependencies across varying time scales. However, the advent of the Transformer architecture (Vaswani 2017) marked a shift, its ability to train in parallel across time steps led to significant efficiency gains and performance improvements. Despite this, the quadratic complexity of the attention mechanism in Transformers remains a bottleneck for deployment in edge and low-power settings. To address this, newer RNN variants such as the *minGRU* (Feng et al. 2024) have been developed to support parallel training while preserving the constant-time, local-state inference characteristic of classical RNNs. These algorithmic simplifications now also open the door to efficient hardware implementations for edge computing scenarios.

One of the key principles behind the design of efficient machine learning accelerators is the reduction of data movement (Bavikadi et al. 2020). This typically manifests itself in IMC, the collocation of memory and computing elements especially for matrix-vector multiplications, as found for example in the linear projections between neural network layers. In this space, analog and mixed-signal implementations typically outperform digital solutions in terms of energy efficiency, especially in a low-precision regime (Sun et al. 2023). This advantage is, however, often diminished by a high complexity and energy footprint of peripheral circuits. A successful mixed-signal implementation must thus attempt to reduce the frequency and resolution of converting between the analog and digital domains, simplify operations to mostly linear and hardware-amenable arithmetics, and reduce the impact of data movement. RNNs, in particular, often require rather complex state update arithmetics and involve a dense recurrent projection of hidden states within a layer. This induces high bandwidth requirements and is typically not compatible to the low-resolution regime where analog implementations perform best.

Here, we present a streamlined RNN architecture that addresses above challenges. It adopts the diagonal-only recurrent projections of contemporary models (Orvieto et al. 2023, Feng et al. 2024), reduces inter-layer communication by enforcing binary output activations, and reduces the complexity of gated state updates by resorting to simplified internal activation functions. Alongside the resulting model, we also introduce a highly efficient mixed-signal implementation thereof. It relies on switched-capacitor circuits to realize both the matrix-vector multiplications involved in the linear projections between layers and the recurrent state updates. Transitions between the analog and digital domains are reduced to a bare minimum and the design can refrain from continuously biased analog circuits such as voltage buffers. The design only relies on metal-oxide-metal (MOM) capacitors, transmission gates, static random-access memory (SRAM) bitcells, and a simple comparator circuit, thus enabling straight-forward scaling and an optimal transfer across technology nodes.

Both architecture and mixed-signal implementation are the result of a stringent co-design process, which allowed to optimally match circuits to the underlying RNN functions. And vice versa, the simplifications to the network model were directly informed by constraints imposed by the microelectronic implementation. The resulting design, to the best of our knowledge the first switched-capacitor implementation of a contemporary gated RNN architecture, thus enables high performance on temporal sequence classification tasks while promising significant efficiency gains compared to the state of the art, which is dominated by purely digital designs (Conti et al. 2018, Paulin et al. 2021, Chen et al. 2024).

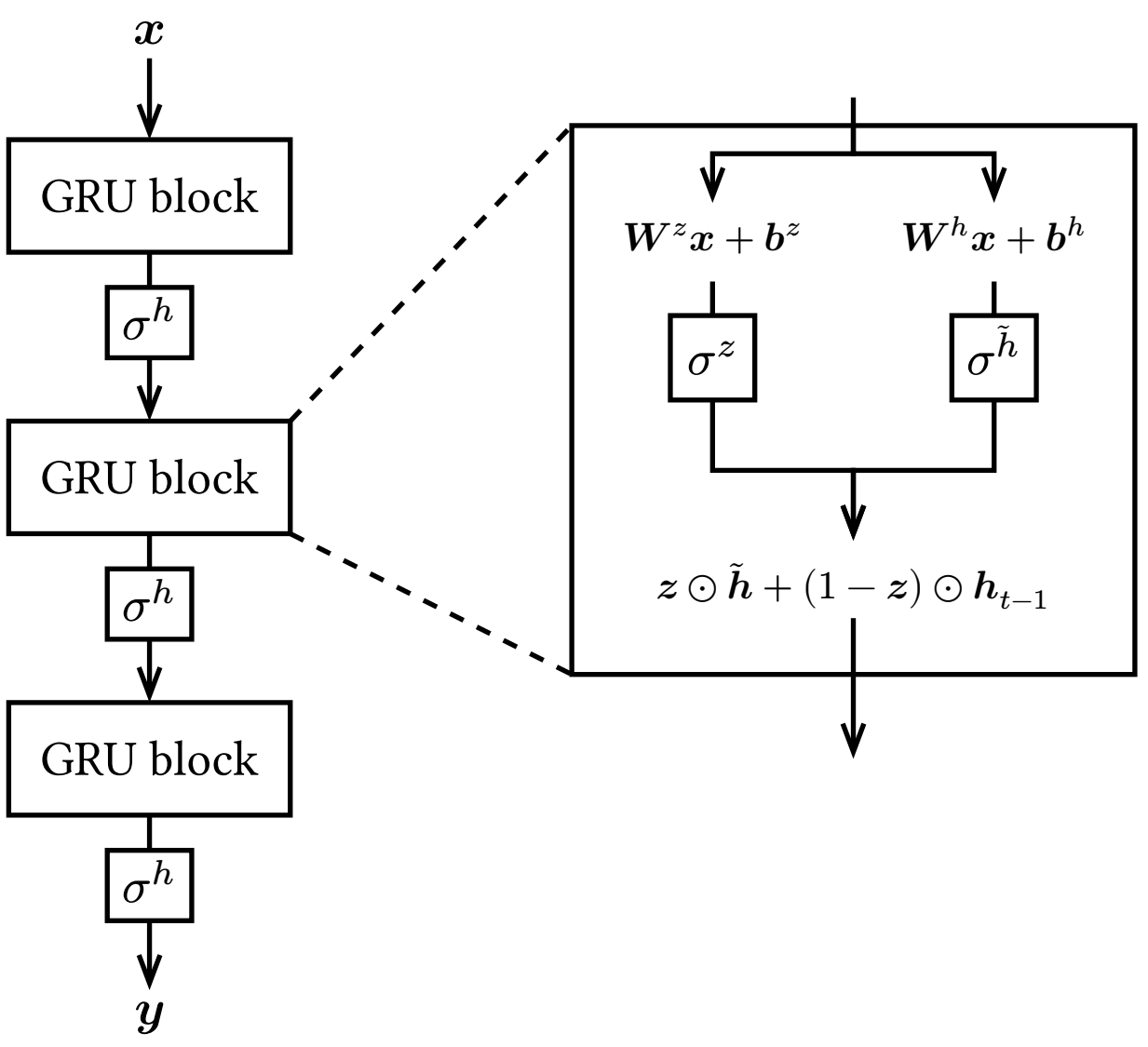

Figure 1: The MINIMALIST architecture consists of a feedforward network of simplified GRU blocks, interleaved with binary activation functions.

## 2 A hardware-amenable GRU-based architecture

Our architecture builds on the *minGRU* model presented by Feng et al. (2024), which describes units with a recurrent internal state

$$h_t = z_t \odot \tilde{h}_t + (1 - z_t) \odot h_{t-1} \tag{1}$$

which persists across time steps $t$ and is only partially overwritten with a *candidate state* $\tilde{h}$ based on the *gate* $z$. The latter two are derived from the units input $x$ through linear projections

$$\tilde{h}_t = W^h \cdot x_t + b^h \,, \tag{2}$$

$$z_t = \sigma^z(W^z \cdot x_t + b^z) \,, \tag{3}$$

with $\sigma$ representing a sigmoidal activation function. In contrast to the original GRUs, *minGRUs* drop the explicit dependency on the previous hidden state $h_{t-1}$ when calculating the gate and proposal states, allowing the application of the highly performant parallel scan algorithm during training. Removing cross-neuron dependencies in the state update – as indicated by the element-wise Hadamard product in Equation 1 – also reduces the recurrent computation to fully local information. This lays an important foundation for an efficient hardware implementation.

Based on the *minGRU* model family, we derive a hardware-amenable architecture. For this purpose, we introduce a number of additional simplifications and constraints.

*Architecture*

To streamline the overall architecture, we refrain from relying on skip-connections or channel-mixing, and instead rely on a simple feed-forward architecture stacking GRU blocks as time mixing units, as shown in Figure 1.

*Quantization*

Most machine learning models rely on floating point numbers to store and compute with their parameters, i.e. neural network weights and biases. Moving to fixed-point or plain integer representations, in contrast, can in many cases dramatically increase the computational efficiency and reduce the memory overhead. Thus, efficient hardware implementations – including most IMC architectures – typically rely on quantized weights for parameter storage as well as computation (Verma et al. 2019).

IMC implementations often push this to the extreme by resorting to low precision representations and even binary weight parameters. This allows them to reduce circuit complexity and optimized resource utilization. Higher precision can often be recovered through time-multiplexing or the aggregation of weight – or "synapse" – circuits into larger units with a then increased overall precision (Verma et al. 2019).

Considering our mixed-signal implementation, we opt for a drastic reduction in weight and bias resolution. Weights are thus quantized to 2 b and biases to 6 b values. In the proposed system, the internal states are represented as analog voltages, and they thus remain unquantized.

*Binary output activations*

We adopt binary output activations to reduce the communication bandwidth between layers – and to simplify the multiply-operation in the linear projections between GRU blocks. To that end, we rely on a Heaviside step function for the output activation function, i.e.,

$$\sigma^h(h_t) = \Theta(h_t) \,. \tag{4}$$

These binary activations allow a sparse, event-based communication of "on" and "off" transitions between layers, in turn reducing the routing fabric's complexity and energy footprint.

*Simplified gating*

To avoid costly arithmetics, we replace the sigmoidal activation function resulting in $z$ by a hard sigmoid, i.e. a piece-wise linear function

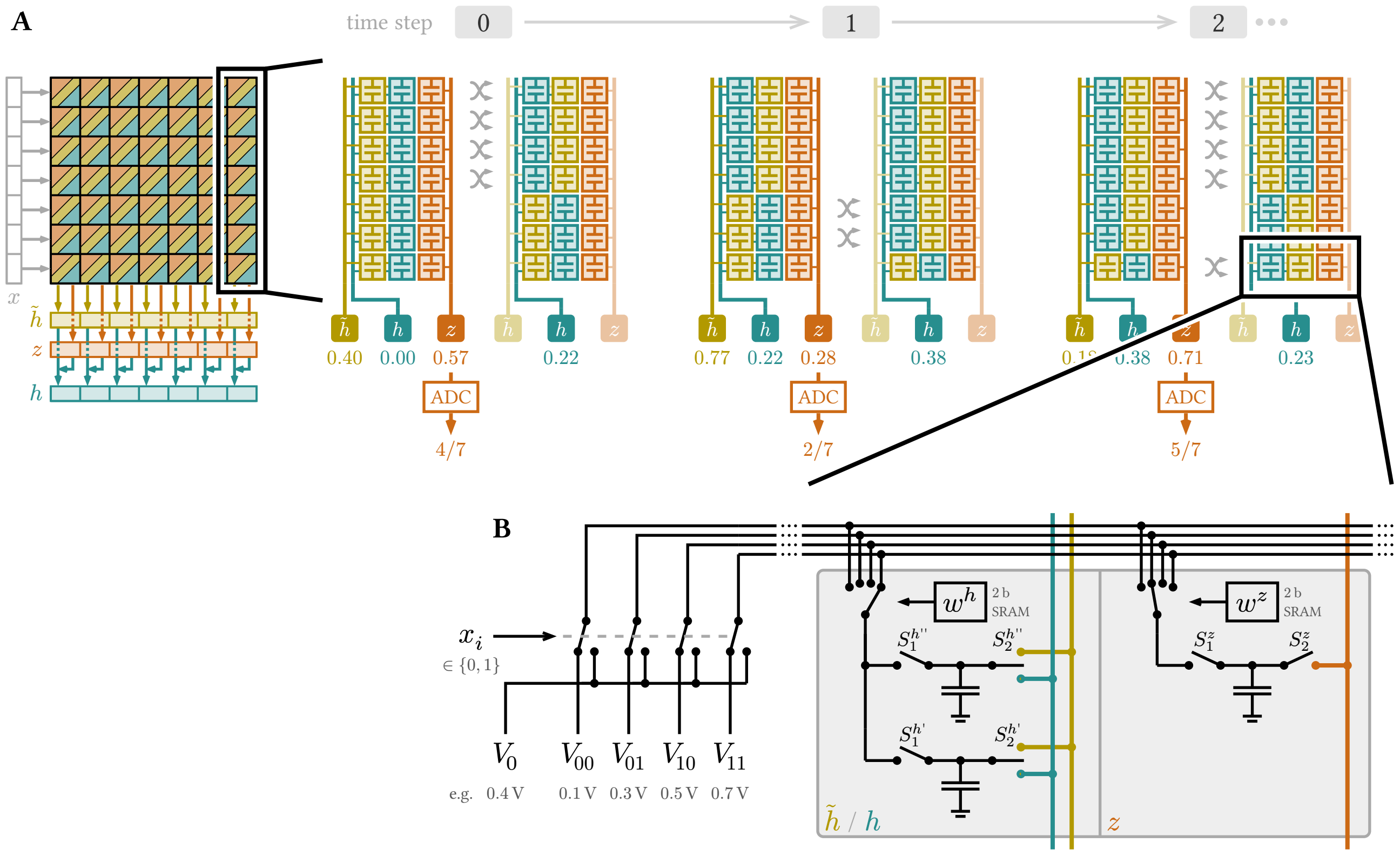

Figure 2: The MINIMALIST cores perform both IMC operations and recurrent state updates through switched capacitor circuits. **A** The cores interleave the weights for the gating ($z$) and hidden state candidates ($\tilde{h}$). Each synapse features three capacitors: one to represent the hidden state $h$, a second one to calculate $\tilde{h}$, and a third for calculating $z$. The first two capacitors swap their roles according to the value of $z$, as shown for exemplary activations across three time steps. **B** More detailed schematic of the switching scheme, including the circuits representing the input activation $x_i$ of a row, represented by a binary value. Weights $w^h$ and $w^z$ are stored in local 2 b SRAM cells, which determine the potential to sample from through turning on one of the 4 switches.

$$\sigma^z(x) = \begin{cases} 0 & \text{if } x \leq -3, \\ 1 & \text{if } x \geq +3, \\ \frac{x}{6} + \frac{1}{2} & \text{otherwise.} \end{cases} \tag{5}$$

The result is then quantized to 6 b.

## 3 Circuit implementation

The MINIMALIST architecture encompasses multiple stacked GRU layers, each connected through feedforward projections. Depending on their dimensionality, these GRU blocks can be mapped to one or multiple cores, which are connected through an event-based routing fabric. The following paragraphs introduce the design of those cores, namely the switched-capacitor IMC and state update circuitry.

### 3.1 Mixed-signal computing cores

The computing cores capture the functionality of a GRU block and the subsequent application of the output activation function $\sigma^h$. To that end, they first calculate the gating variable $z$ as well as the new proposal state $\tilde{h}$ that both result from linear input projections through IMC. They also implement the subsequent state update mechanism.

#### *3.1.1 Switched-capacitor-based IMC input projections*

MINIMALIST realizes the linear input projections representing $\boldsymbol{W}^h$ and $\boldsymbol{W}^z$ through switched-capacitor IMC matrices. The two resulting matrix-vector multiplications share the same input vector $x$, and can thus be merged into a single matrix as indicated in Figure 2A. Each GRU circuit is thus connected to a column of $h$ and $z$ synapses, each.

An $h$ synapse fulfills two distinct roles: it participates in the respective matrix-vector multiplication, but is also responsible for maintaining the previous hidden state $h_{t-1}$. It thus features two identical capacitors. At each point in time, one of them holds the previous hidden state $h$ and is involved in the state update calculations (ref. Section 3.1.3). The second is available to calculate the new candidate state $\tilde{h}$ through IMC (Figure 2B).

In general, capacitor-based IMC solutions profit from the comparably accurate matching of metal-based capacitor structures – even when relying only on parasitic fringe

capacitances – to achieve state-of-the-art accuracy at a minimal energy budget and a compact silicon footprint (Bankman et al. 2018, Valavi et al. 2019). Capacitive IMC can be achieved through either charge redistribution or charge sharing strategies. Both are compatible with multi-bit multiply-accumulate (MAC) operations and, for that purpose, typically rely on segmented capacitors for a fractional control over the charge and thus the effective weight. As the presented architecture relies on the sampling capacitors also for maintaining and updating of the internal GRU states $h_t$, the sampling nodes must possess a constant and known capacitance. Multi-bit weights thus have to rely on multiple voltage levels to modulate the charge.

We opted for a charge sharing paradigm and realized a 2 b weight resolution by allowing each synapse to choose among four distinct, equidistant voltages, $V_w, w \in \{00, 01, 10, 11\}$, based on the locally stored weight (2b SRAM cell, refer to Figure 2B). $V_0 = \frac{1}{2}(V_{00} + V_{11})$, a fifth potential representing zero activations, is chosen at an intermediate voltage, thus resulting in two positive and two negative weight values, although this might be adapted also on a per-layer basis to better represent the weight statistics of a given network.

Calculating the element-wise product of binary input activation $x_i$ and weight $w_{ji}$ at the intersection of row $i$ and column $j$ involves both the synapse itself and the row-wise driver circuitry (Figure 2B): Presenting $x_i = 1$ connects the four shared horizontal lines to the weight potentials $V_w$. In case of $x_i = 0$, they are clamped to $V_0$. The synapse then choses to sample from one of those four lines according to the locally stored weights.

To compute both $z$ and $\tilde{h}$, the respective sampling capacitors are first pre-charged via $S_1^*$ to the weight potentials corresponding to the weight values stored in local SRAM. In a second phase, the capacitors within a column are shorted via switches $S_2^*$. As a result, they share their charge and the potential settles towards

$$\boldsymbol{V}^{z,\tilde{h}} = V_w\left(\boldsymbol{W}^{z,\tilde{h}}\right) \cdot \boldsymbol{x}_t \cdot \frac{1}{\dim(\boldsymbol{x}_t)}, \qquad (6)$$

representing the means of the weighted input activations and thus implementing the desired linear projections.

*3.1.2 Digitization of z*

The computation of state updates further relies on switching of capacitors based on the value of $V^z$ (ref. Section 3.1.3). To that end, $z$ has to be known in the digital domain, which we address with a 6 b successive-approximation register (SAR) analog-to-digital converter (ADC) (Figure 3).

We can directly apply the activation function $\sigma^z$, a hard sigmoid, by restricting the dynamic range of the ADC. A limited range, typically caused by parasitic capacitances reducing the voltage swing of the ADC's capacitor array, poses a challenge in many applications. Here, we deliberately exploit this effect by keeping the sampling capacitors $C_{\mathbb{1}}^z$ connected during the digitization phase (Figure 3A). Segmenting the IMC matrix into groups with a binary scaling enables granular control over switches $S_2^z$. This allows to disconnect parts of the IMC sampling capacitors after charge sharing, inducing control over the ratio $C_{\mathrm{ADC}}/C_{\mathrm{IMC}}$ and thus the ADC's dynamic range (Figure 3B). Therefore, the circuits can be ideally matched to the layer-specific slope of $\sigma^z$ (Figure 3 C).

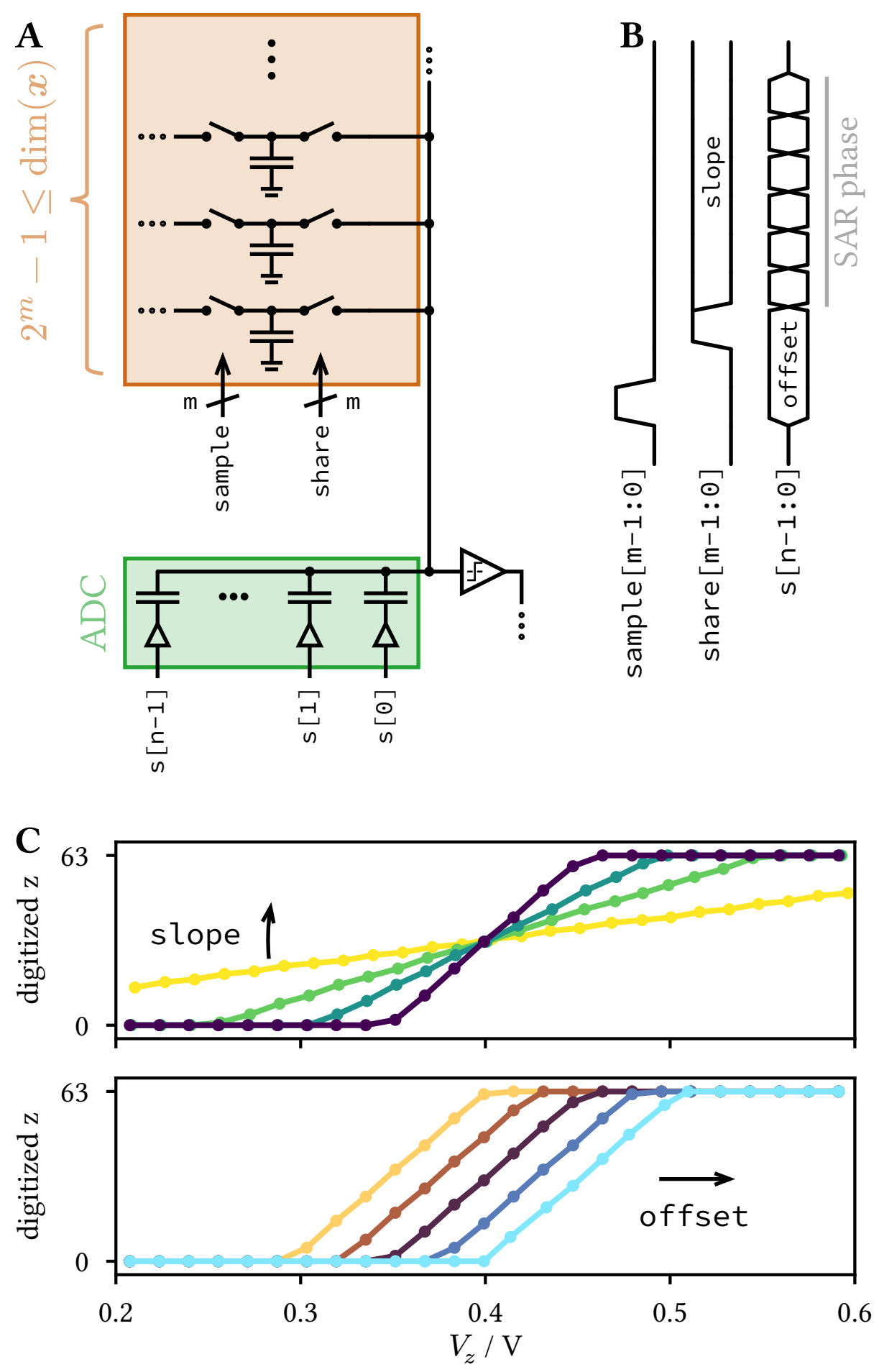

Figure 3: The ADC transfer characteristics can be tuned by controlling the capacitive load represented by the IMC array or by using the capacitive DAC to induce a constant bias. This allows mapping of a wide range of $\sigma^z$ activation functions, also on a per-layer or per-unit basis. **A** Schematic of a SAR ADC channel and a column of $z$ synapses. The sharing switches can be controlled to connect or disconnect a variable number of synapses to the ADC's input, thus allowing to tune the ratio of capacitances and in turn the slope of the activation function. **B** Timing of the sampling and sharing phases calculating $z$, but also of the digitization through successive approximation and, importantly, the pre-charging of the ADC's capacitor array to induce an offset. **C** Mixed-signal simulation results showing ADC characteristics as a function of the `slope` and `offset` parameters.

For a constant bias on $z$, we can rely on the ADC's capacitive digital-to-analog converter (DAC) to generate an offset on the sampled potential: During the sampling phase, the capacitor array is pre-set to a 6 b `offset` before then starting the successive approximation with the initial configuration (`s[5:0] = 0b100000`). This allows shifting of the ADC's transfer characteristics by half of the dynamic range towards both positive and negative voltages (Figure 3C).

#### 3.1.3 State update through charge sharing

The state update itself is, again, implemented through charge sharing. Within a column, each synapse contributes one sampling capacitor to represent the previous hidden state $h_{t-1}$ represented as $V^h$ on the total capacitance $C^h$, while the other capacitor is used to calculate $\tilde{h}$ according to the IMC scheme introduced before, now present as $V^{\tilde{h}}$ on the merged capacitance $C^{\tilde{h}}$. Updating $h$ simply involves mixing the charge between $C^h$ and $C^{\tilde{h}}$ with a weighting determined by $z$, the 6 b digital representation of $\sigma^z(V^z)$. For that purpose, the circuits, again, rely on a segmented IMC matrix and thus granular control over $S_2^h$.

The number of swapped capacitors is simply proportional to $z$. When $z = 0$, the capacitor bank representing $h$ remains untouched, when $z = 1$, all capacitors are exchanged and thus fully carry over $\tilde{h}$. Intermediate values of $z$ result in a proportional mixing of the two state variables. The process of updating $h$ is thus equivalent to swapping sampling capacitors between the two output lines $h$ and $\tilde{h}$, effectively reassigning their role to either represent the previous hidden state or to be available for calculating the next $\tilde{h}$ (Figure 2A). Crucially, this scheme does not require buffering of the internal states and simply redistributes charge between the capacitors. This reduced the overall energy footprint but also design complexity by restricting itself only to capacitors and switches.

#### 3.1.4 Output activations

The GRUs' output activations, represented by the Heaviside step function, are applied by reusing the ADC's comparator circuit. A bias on $h$ can be subsumed in the comparator's reference potential, which is generated through the ADC's capacitive DAC.

### 3.2 Implementation

For the IMC arrays and the state-update circuits, the full-custom computing cores only rely on commodity circuits, such as transmission gates, SRAM bitcells, transmission gates, and a simple comparator used in the ADC design. This allows an optimal scaling and transfer across technology nodes. We opted to implement the MINIMALIST architecture in Globalfoundries' 22 nm FD-SOI process, and can fully rely on their dense core transistor offerings.

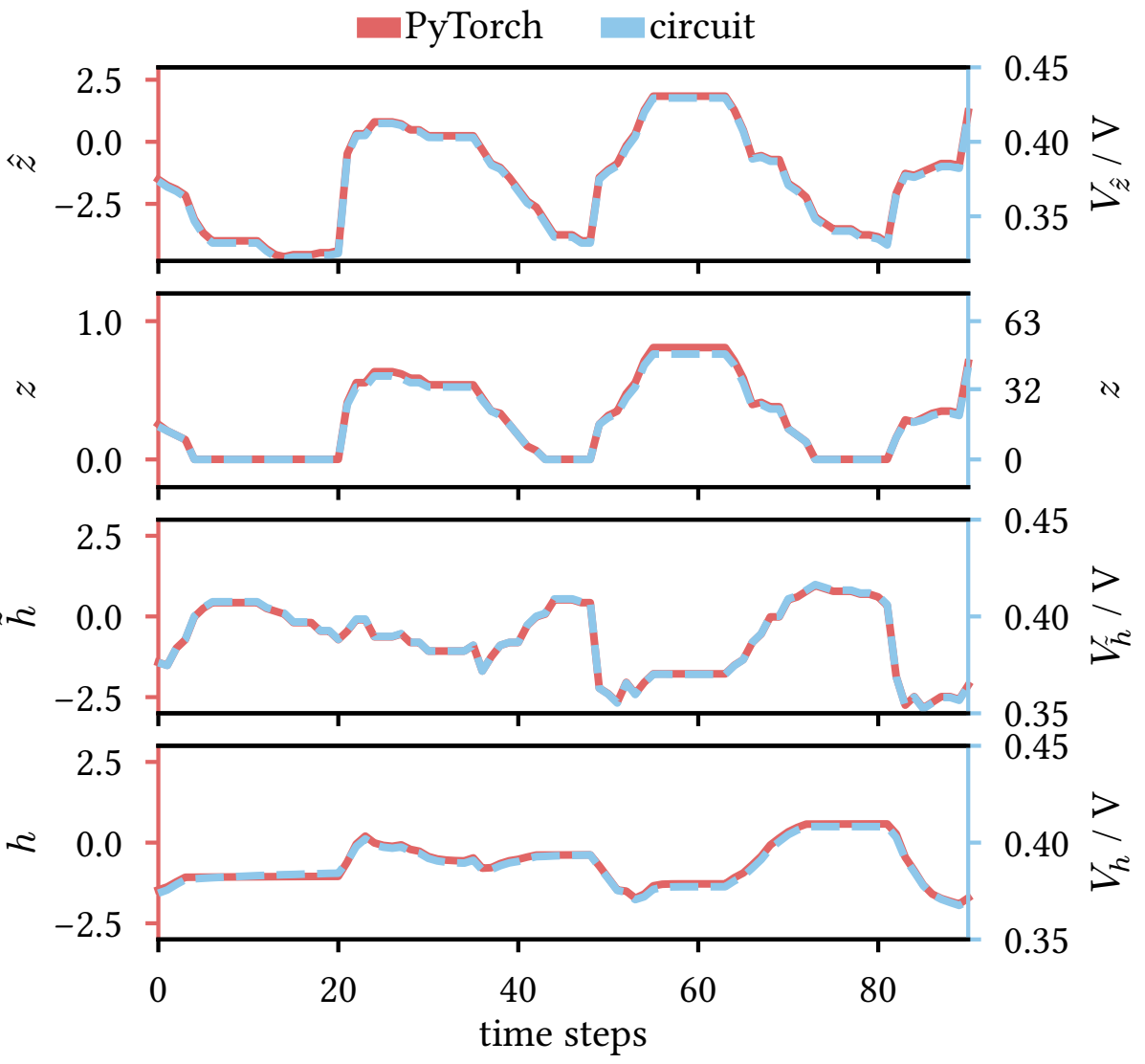

Figure 4: Comparison of activations recorded from a software implementation of the model and a mixed-signal simulation set up with equivalent weights and biases. The traces stem from a random unit within a network trained on the sequential MNIST dataset.

## 4 Results

We verified the MINIMALIST circuits in mixed-signal simulations using Cadence Spectre AMS Designer. To that end, we extracted weights, biases, and input activations from a model implemented and trained in PyTorch, and the circuit simulation was set up accordingly. Figure 4 compares the resulting activations on $z$, $\tilde{h}$, and $h$ between the original software and the circuit implementation.

### 4.1 Network performance

Additionally, we have evaluated the performance impact of quantization and adaptations to the network architectures necessary for hardware deployment. In Figure 5 we compare the performance of three networks on the sequential MNIST dataset. All three share the same number of layers and GRU-blocks per layer (1-64-64-64-64-10) and with that include the same number of trainable parameters. The baseline network is trained in full 32 b floating-point precision, uses the same activation functions as described in the original publication (Feng et al. 2024) and achieves a test accuracy of 98.1 %. When quantizing the weights to 2 b integers, the biases to 6 b and binarizing $\sigma^h$, while keep-ing the internal GRU-states and activation functions the same, we incur a performance penalty of 0.4 % – at a ten-fold reduction of parameter memory. This, however, requires the extension of the network training to a multi-stage process of 4 gradual phases of quantization-aware training. For hardware compatibility it is additionally required to eliminate the actiavation function on $\sigma^{\tilde{h}}$, exchange $\sigma^z$ with a hard-sigmoid and to quantize the gating variable $z$ to 6 b integers. With this the network reaches a final test accuracy of 96.9 %.

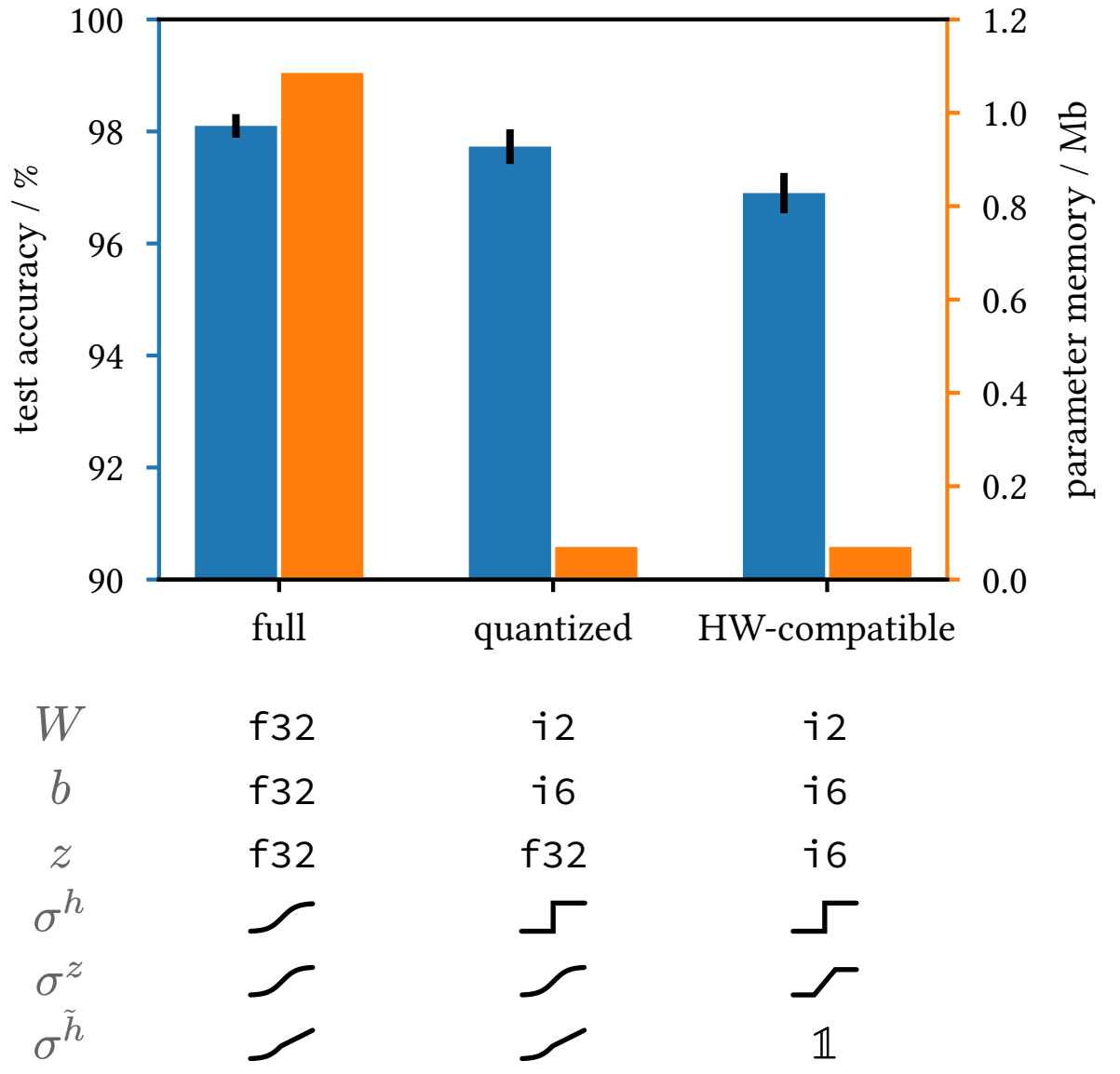

Figure 5: Performance of our model on the sequential MNIST dataset (mean and standard deviation across 10 seeds). We trained three models: The first one relied on the original activation functions and a floating point representation of weights and biases as well as internal activations. The second was restricted to quantized weights and biases as well as 1 b output activations. The third model was fully compatible to the hardware constraints and to that end, also included a quantized hard sigmoid activation function on $z$.

### 4.2 Energy efficiency

The energy expenditure of the mixed-signal computing cores is dominated by the repeated charging and discharging of the sampling capacitors, as well as the toggling of the switches. Considering a network spanning 4 cores with 64 rows and 64 columns each, we estimate the energy to be bounded by 169 pJ per time step. Here, we assume all switches to toggle – the worst case scenario corresponding to a constant $z = 1$. Our estimate, however, does not yet include the SAR ADC (with a total DAC capacitance far below the IMC capacitance), the event routing (with relatively sparse 1 b activations), the digital control logic, and clock distribution.

## 5 Discussion

In this manuscript, we have introduced the MINIMALIST architecture based on simplified GRUs. The network incorporates constraints on its weight quantization and activation functions to make it amenable for a mixed-signal circuit implementation. The implementation relies on switched-capacitor circuits to implement the linear weight projections as well as recurrent state update circuit, and is free of continuously biased analog blocks. Besides the comparator used within the SAR ADC, it only consists of SRAM bitcells, transmission gates, and MOM capacitors, and can thus be optimally scaled and transferred across technology nodes. To the best of our knowledge, the presented architecture and circuits represent the first switched-capacitor implementation of a contemporary RNN architecture.

We have verified the architecture on the sequential MNIST dataset and compared the circuit dynamics to data recorded from the software model in a mixed-signal simulation. Tentative energy estimates promise significant efficiency gains compared to the state of the art (Giraldo and Verhelst 2018, Ankit et al. 2019, Zhao et al. 2019), but more elaborate estimates and analyses are required for a fair comparison.